# The differential geometry of perceptual similarity

## Antonio M Rodriguez[1] and Richard Granger[1]*

[1]Brain Engineering Lab, Dartmouth College, Hanover NH 03755, USA
*Richard.Granger@gmail.com

## Abstract

Human similarity judgments are inconsistent with Euclidean, Hamming, Mahalanobis, and the majority of measures used in the extensive literatures on similarity and dissimilarity. From intrinsic properties of brain circuitry, we derive principles of perceptual metrics, showing their conformance to Riemannian geometry. As a demonstration of their utility, the perceptual metrics are shown to outperform JPEG compression. Unlike machine-learning approaches, the outperformance uses no statistics, and no learning. Beyond the incidental application to compression, the metrics offer broad explanatory accounts of empirical perceptual findings such as Tversky's triangle inequality violations (1, 2), contradictory human judgments of identical stimuli such as speech sounds, and a broad range of other phenomena on percepts and concepts that may initially appear unrelated. The findings constitute a set of fundamental principles underlying perceptual similarity.

## Introduction: The fundamentals of perceptual similarity

When do images look alike? All standard Euclidean (and Hamming, and Mahalanobis, and almost all other) standard measures of similarity turn out to be at odds with human similarity judgments. We spell out why this is the case, give explanatory principles, and provide an illustrative application to the widely-used JPEG compression algorithm: JPEG has been outperformed via extensive learning by neural network and ML methods, whereas we outperform it with no statistics, and no training. We show that the JPEG findings fall out as a special case of the underlying broad principles introduced here, which are applicable to a wide range of unsupervised methods that entail similarity measures.

Euclidean vectors' components are orthogonal, and thus $\vec{a} = (10000)$ and $\vec{b} = (00010)$ are equidistant from $\vec{c} = (00001)$: distances $\|\vec{a}\vec{c}\|$ and $\|\vec{b}\vec{c}\|$ both have Hamming distances of 2, and Euclidean distances of $\sqrt{2}$. However, considered as physical images, the right-hand positioning of the "1" values in vectors $\vec{b}$ and $\vec{c}$ render them more visually similar to each other than either is to $\vec{a}$. When such "neighbor" relations within a vector are considered, then vector axes are not orthogonal, and non-Euclidean metrics can readily yield smaller distances between $\vec{b}$ and $\vec{c}$ than between $\vec{a}$ and $\vec{c}$.

Human judgments of similarity imply a particular geometric system, and as in the above simple example, it is easy to show that human similarity judgments do not conform to Euclidean, Hamming, Mahalanobis, or the other most commonly used similarity metrics; rather, we will show that they conform to Riemannian geometry.

Moreover, human similarity judgments are not solely a function of the stimuli themselves; they depend also on the operations internally carried out by the perceiver. Given physical stimuli (e.g., the English speech sounds /ra/ and /la/) are distinguishable to some perceivers (native English speakers) but difficult to discriminate by other perceivers (e.g., native Japanese speakers), as a result of prior experience of the perceiver (3-7). Thus "re-coding" the stimuli does not contribute to a solution (8-10). Rather, we wish to identify the perceptual metrics that a given perceiver uses when judging similarity and difference.

As in the vector "neighbor" example, perceptual brain system anatomy reflects non-Euclidean metrics: the signal transmitted from one group of neurons to another directly corresponds to mappings among non-Euclidean spaces. We derive a formalism from synaptic connectivity patterns and show that the system matches and explains individual human empirical judgments.

We apply the derived system to the well-studied JPEG compression task (solely to demonstrate the real-world efficacy of the presented principles).

Conveniently, several statistical machine-learning approaches have recently been shown to outperform the sturdy hand-constructed JPEG method, via extensive training on image data (11, 12). Those results may be viewed as calling for potential explanatory principles that may underlie their successes. What structure in the data is being statistically identified by these learning approaches?

The method derived in the present manuscript is easily shown capable of outperforming JPEG as well, with no increase in computational cost over JPEG, and using no statistics and no training. The results arise from newly posited principles that underlie not just JPEG, but perceptual similarity in general; JPEG is shown to fall out as a special case of the method.

Beyond the illustrative JPEG example, the metrics are shown to proffer explanatory accounts of a range of empirical perceptual findings, notably Tversky's triangle inequality violations (1, 2), contradictory human judgments of stimuli in other modalities such as speech sounds, and beyond simple percepts to abstract concepts and categories that may initially appear unrelated.

## Results
### What JPEG does, and a principled error
Any lossy compression algorithm trades off image quality for image entropy: how the image appears vs. how much space it can be stored in. The JPEG standard (13, 14) transforms an image into a frequency basis, and encodes each of the frequency components with a different amount of precision (tending to encode low-frequency components with more precision than high-frequency components), thus selectively introducing modest errors preferentially into high frequency components, yielding a new image with more error but less entropy than the original; i.e., an image judged (via human viewers) to be of lesser quality, but capable of fitting into a smaller file size.

An axiomatic error incorporated into JPEG (and indeed into most compression methods, and most measures of similarity and dissimilarity), is the assumption that the frequency basis vectors are orthogonal, and thus that changes to any one of them do not impact the others. In Euclidean space, these bases are indeed orthogonal; in Riemannian space they are not. We show that human perceptual similarity judgments are consistent with non-orthogonal bases, properly treated as a Riemannian space, not Euclidean or affine.

For JPEG, let $\bar{p}$ be a 64-dimensional vector comprised of intensity information from an 8x8 block of pixels (we initially focus on monochrome images for expository simplicity). We define a matrix $F_{x,y}$ whose elements are the intensities of the pixels at locations $(x,y)$. JPEG encoding proceeds via the following four steps:

1) Center image intensities around 0: $F'_{x,y} = F_{x,y} - 128$.

2) Represent image as linear combination of frequency components
   The discrete cosine transform (DCT) operator $T$ is given by:
   $$T_{u,v} = \tfrac{1}{4}\alpha(u)\alpha(v)\sum_{x=0}^{N-1}\sum_{y=0}^{N-1} F'_{x,y} \cos\left[\frac{(2x+1)u\pi}{2N}\right]\cos\left[\frac{(2y+1)v\pi}{2N}\right] \quad (2.1)$$
   where $\alpha(x) = \begin{cases} 1/\sqrt{2} & \text{if } x=0 \\ 1 & \text{else} \end{cases}$ and where N=8 for JPEG.

   (The DCT is a linear operator acting on the (centered) image $F'$; i.e., $T(F') = T \bullet F'$)

3) Quantization and rounding of frequency components
   Each of the 64 dimensions of T are independently scaled by a predetermined "quantization" matrix $Q_C$ (with the intended effect of discarding less-relevant information in the data), with different matrices defined for different "calibers" $C$ of the error/entropy tradeoff. The operator $R$ simply divides the elements of a transformed input ($T$) by the designated quantization matrix $Q_C$: $R_{u,v} = T_{u,v} / Q_{C_{u,v}}$ or, in matrix notation, $R = Q_C^{-1} \bullet T$

   Full quantization of input $F'$, then, is accomplished in JPEG by
   $$Z = \text{round}(R \bullet F') = \text{round}(Q_C^{-1} \bullet T \bullet F') \quad (2.2)$$
   The JPEG standard establishes preëstablished quantization matrices $(Q_C)$ for any given desired compression factor $C$, i.e., for a given reduction in quality and commensurate decrease in entropy. These quantization matrices were constructed by hand (13, 14), with no non-manual method for arriving at its values (15).

4) Entropy encoding
   The numerical elements of JPEG's quantized and rounded image $Z$ are encoded via Huffman coding, such that the most frequently used numerical values are assigned the shortest bit representation, thus taking advantage of the reduced entropy of the quantized input, to enable the compressed image to be stored in a file smaller than the original. Although Huffman coding is used by JPEG, any number of entropy encoding methods (such as arithmetic encoding) would suffice.

We focus on a specific erroneous assumption underlying JPEG (and other perceptual compression methods): the use of Euclidean measures of similarity. In fact, any given pixel ***p*** is not perceptually independent of nor orthogonal to neighboring pixels. As in the examples in the introduction section, neighboring pixels are perceived by a viewer as being "closer" to each other than are more distal pixels, and this alters perceived similarity.

Correspondingly, the 64 frequency bases in the DCT are not perceptually orthogonal (though they are orthogonal in Euclidean space): some are perceptual judged more similar to each other than others by human viewers. Moreover, these judgments are dependent on their coefficients, and thus have different similarities in different parts of the DCT basis space. The curvature of the space thus is not affine, but rather Riemannian (Figure 1).

We introduce an appropriate Riemannian treatment of perceptual similarity. We show that the resulting method can readily outperform JPEG, but more importantly, it has explanatory power: JPEG emerges as a special case of the general method, and the underlying geometric principles of human perception become more closely explained.

**The Riemannian geometric principles of perception**
*Three geometric spaces*
Assume an image of 64 pixels (grayscale, for temporary pedagogical simplicity), arranged as an 8x8 array. In JPEG and all other standard compression mechanisms, the image is treated as an arbitrarily, but consistently, ordered 64-dimensional vector, such that each vector entry corresponds to the intensity at one of the 64 pixels in the 8x8 array. This renders the data into simple vector format, enabling the applicability of vector and matrix operations. However, it does so at the cost of eliminating the neighbor relations among pixels in the physical space. (Typically, the 64 pixels are ordered (arbitrarily) with entries 1-8 from the top row of the array, 9-16 from the next row, and so on.) The elimination of neighbor relations would be irrelevant if human perception of a pixel were modulated equally, or not at all, by the characteristics of neighboring and distant pixels alike; this turns out not to be the case. We forward the alternative in which the image is described in terms of a physical space $\Phi$ with 3 dimensions (for x and y locations, and intensity), for each of the 64 pixels. This corresponds to a transformation of the "feature" space $\mathcal{F}$ into physical space $\Phi$. The explicit representation of physical pixel positions enables perceptual encodings that use pixel position as a parameter.

In addition to feature space $\mathcal{F}$ and physical space $\Phi$, we introduce a third space, $\Psi$, which we term "perceptual space," that includes representation of perceptual geometric relations among the image elements (Figure 2). Transforms into this perceptual space, accomplished by differential geometry, will be shown to directly correspond to human perceptual similarity judgments.

*Brain connectomes are Riemannian*
Figure 2 shows sample anatomical connectivity among brain regions, and its formal properties. Figure 2a shows an instance of typical mammalian thalamocortical and cortico-cortical synaptic projections (16-20). The projection pattern from one cellular assembly to another is not perfectly "point to point" (i.e., each cell projecting to exactly one topographically corresponding target cell) nor completely diffuse (with no topography); rather, the projection is "radially extended," such that each element contacts a range of targets roughly within a spatial neighborhood or radius around a target. Figure 2b shows a simple vector encoding of these projection patterns with corresponding synaptic weights $n', n''$, etc. Figure 2c shows examples of typical physiological neural responses in early sensory cortical areas that can arise from these connectivity patterns. Figure 2d contains the general form of a Jacobian matrix denoting the overall effect of activity in the neurons of an input area $x$ on the neurons in target area $f$; each entry in the Jacobian designates the change in an element of $f$ as a consequence of a given change in an element of $x$. Figure 2e is an example instance of such a Jacobian, corresponding to the synaptic connectivity pattern in Figure 2b.

Intuitively, a Jacobian encodes the interactions among stimulus components. If a vector contained purely independent entries (as in an imagined perfectly point-to-point topography with no lateral fan-in or fan-out projections), the Jacobian would be the identity matrix: ones along the diagonal and all other entries zeros. Each vector dimension then has no effect on other dimensions: a given input unit affects only a single target unit, and no others. In actual connectivity, which does contain some radially extended projections, there are off-diagonal non-zero values in the Jacobian corresponding to the slightly non-topographic synaptic contacts (Figure 2b).

When the input and output patterns are treated as vectors, any off-diagonal Jacobian elements reflect influences of one dimension on others: the dimensions are not orthogonal, and the vectors are Riemannian, not Euclidean (21).

All perceptual systems can be seen to intrinsically express a "stance" on the geometric relations that occur among the components of the stimuli processed by the system. In the degenerate case of no off-diagonal elements, the system would act as though it assumes independence of components (Euclidean vectors). In all pathways characteristic of most thalamo-cortical and cortico-cortical projections, however, the processing inherently assumes non-Euclidean neighbor relations among the stimuli.

It is notable that any bank of neuronal elements with receptive fields consisting either of Gaussians or of first or second derivatives of Gaussians, will have precisely the effect of computing the derivatives of the inputs in just the form that arises in a Jacobian (see equation (2.3) below) (22-26). Physiological neuron response patterns thus appear thoroughly suited to producing transforms into spaces with Riemannian curvatures. (Notably, this implies that a synaptic change (e.g., LTP) causes specific re-shaping of neurons' receptive fields, modifying the curvature of the space of the target cells in a given projection pathway.)

The matrix ***J*** in Figure 2d describes the particular transform from an input space to an output space. This is an instance of specifying the differences between a perceptual input, versus a percept that is received via this projection pathway. A perceiver will process an input as though it contains the neighbor relations specified by the Jacobian.

<u>*The map from physical to perceptual space*</u>
Neighboring entries in a vector, like adjacent notes on a piano keyboard, are closer to each other than entries from other, non-neighboring dimensions. The features thus do not constitute independent dimensions (or, put differently, the dimensions are not orthogonal). In these (extremely common) cases, target perceptual distances are correctly rendered by Riemannian rather than by Euclidean measures. It is notable that this not an exception but the normal case for perception. Euclidean vectors do not treat constituents as having neighbors, but perceivers do.

We wish to determine, then, how changes to an image will be perceived. A change in physical space (i.e., the image) can be directly measured. The corresponding predicted change in perceptual space can then be computed via a metric tensor which measures distance in the perceptual space with respect to positions in physical space. This metric tensor is computed via a Jacobian that maps from distances in physical space to distances in perceptual space (Figure 3). For the map $\mu_{\mathcal{F} \to \Phi}$ the Jacobian $J_{\mathcal{F} \to \Phi}$ will be a 3x64 matrix:

$$J_{\mathcal{F} \to \Phi} = \begin{bmatrix} \frac{d\Phi_0}{df_0} & \frac{d\Phi_0}{df_1} & \frac{d\Phi_0}{df_2} & \cdots & \frac{d\Phi_0}{df_{63}} \\ \frac{d\Phi_1}{df_0} & \frac{d\Phi_1}{df_1} & \frac{d\Phi_1}{df_2} & \cdots & \frac{d\Phi_1}{df_{63}} \\ \frac{d\Phi_2}{df_0} & \frac{d\Phi_2}{df_1} & \frac{d\Phi_2}{df_2} & \cdots & \frac{d\Phi_2}{df_{63}} \end{bmatrix} \quad (2.3)$$

(Supplemental sections §2.5-§2.11 give sample values used to generate specific Jacobians for image processing, as in the examples shown in Figure 4).

This Jacobian enables identification of a distance metric for the feature space $\mathcal{F}$ with respect to its embedding in physical space $\Phi$ in terms of the metric tensor **g** (see Supplemental section §2.5 for examples of values used):

$$\mathbf{g}_{\Phi:\mathcal{F}}(\vec{x}) = J^T_{\mathcal{F} \to \Phi}(\vec{x}) \bullet J_{\mathcal{F} \to \Phi}(\vec{x}) \quad (2.4)$$

i.e., the metric tensor **g** is using measures in space $\Phi$ applied to objects in space $\mathcal{F}$, or, put differently, the tensor measures distances in $\mathcal{F}$ with respect to measures in space $\Phi$. Then, mapping physical space $\Phi$ to perceptual space $\Psi$ (via Jacobian operator $J_{\Phi \to \Psi}$) defines how features in the physical space are perceived by a viewer, enabling a formal description of how changes in the physical image are registered as perceptual changes.

*Specific construction of the Jacobian mapping* $J_{\Phi \to \Psi}$

Information from the physical stimulus or from the perceiver (or both) enables construction of a Jacobian to map from physical vectors to the perceptual space a perceiver may use. Such a Jacobian, $J_{\Phi \to \Psi}$, can be obtained directly from either synaptic connectivity patterns or from psychophysics – either by a priori assumptions or from empirical measurements.

(i) Synaptic Jacobian:
   a) *Empirical*
      Measure anatomical connections and synaptic strengths, if known; the Jacobian is directly obtained from those data as in Figure 2. These measures typically are unavailable, but as will be seen, approximations may be drawn from a set of simple connectivity assumptions.
   b) *Estimated*

Assume radius of projection fan-out from a cortical region to a target region (Figure 2a), based on measures of typical such projections in the literature (16-20), and estimate a factor by which distances among stimulus input features (e.g., pixels) influence relatedness of the features, and resulting curvature of the Riemannian space in which they are thus assumed to be perceptually embedded.

(ii) Psychophysical Jacobian:

  a) *Empirical*

  Measure constituent physical features of the stimuli and calculate distances among stimulus features, such as pixel size, pixel disparity, viewing distance, and obtain empirical measures of human-reported distances; the Jacobian is the set of relations among psychological and physical distances as in Equation (2.3).

  b) *Estimated*

  Assume Gaussian fall-off of relatedness of neighboring pixels in a stimulus; measure constituent features as in (ii a) and estimate a factor by which distances between stimulus input features (e.g., pixel distances in x and y directions) influence the relatedness of the features (and the resulting curvature of the Riemannian space in which they are assumed to be perceptually embedded).

In the present work we proceed with method (ii b), i.e., measuring (Euclidean) physical distances among pairs of inputs and positing a range of candidate factors by which the physical disparity among features may give rise to perceived feature interactions. We show a series of resulting findings corresponding to this range of different hypothesized factors (Supplemental section §2.5).

Having obtained a Jacobian by any of the above means, we compute metric tensor **g** as in equation (2.4). (The tensor alternately may be obtained in condition (ii) using the covariance matrix $\Sigma$ from psychophysical experimental data: $\mathbf{g}_{\Psi:\Phi} = \Sigma_{\Psi:\Phi}^{-1}$.) (See supplemental section §1.4).

We may move the obtained metric $\mathbf{g}_{\Psi:\Phi}$ from physical space to feature space, obtaining a new metric $\mathbf{g}_{\Psi:\mathcal{F}}$ that measures distances in the feature space with respect to the perceptual space:

$$\mathbf{g}_{\Psi:\mathcal{F}} = J_{\mathcal{F}\to\Phi}^T \cdot \mathbf{g}_{\Psi:\Phi} \cdot J_{\mathcal{F}\to\Phi} \qquad (2.5)$$

This new metric in the feature space now computes the Riemannian distances among dimensions that hold in the feature space.

The metric can be used to compute the matrix of all distances among all pairs of features $x_i, x_j$ in a column vector $\bar{x}$ of dimensionality $k$:

$$dist(x_i, x_j) = \left((2\pi)^k \left|\mathbf{g}_{\Psi:\mathcal{F}}\right|\right)^{-1/2} \exp\left(-\tfrac{1}{2}(x_i - x_j)^T \mathbf{g}_{\Psi:\mathcal{F}} (x_i - x_j)\right) \qquad (2.6)$$

(where $|\bar{a}|$ is the determinant of $\bar{a}$). (Sample distance matrices for selected specific measured visual parameters are shown in Supplemental section §2.7; tables §2-§4).

The dimensions of a feature vector in (Euclidean) space $\mathcal{F}$ are orthogonal, but the dimensions of the corresponding vector in (Riemannian) perceptual space $\Psi$ are not orthogonal; rather, the pairwise distances among the dimensions are described by Eq. (2.6).

## Methods
### Derivation of Riemannian geometric JPEG (RGPEG)
JPEG modifies the image feature vector, introducing error (the distance between the original and modified vector), such that the modified vector has lower entropy, and thus can be stored with a smaller description.

Correspondingly, we too will modify the feature vector, introducing error in order to lower entropy, but in this case using graph-based operations on non-Euclidean dimension distances (Eq. (2.6)). We introduce the Riemannian geometric perceptual encoding graph (RGPEG) method.

JPEG uses a (hand-constructed) quantization ("$Q$") matrix that specifies the amount by which each of the 64 DCT dimensions will be perturbed, such that when they are subjected to integer rounding, they will exhibit lower entropy.

We replace the JPEG quantization operations with a principled formula that computes perturbations of basis dimensions to achieve a desired entropy reduction and commensurate error – but in perceptual space rather than in feature space. Specifically, the surrogate quantization step moves the image in perceptual space along the gradient of the eigenvectors of the Hamiltonian of the basis space. We show that the resulting computation can outperform JPEG operations (or any operations that take place in feature space rather than in perceptual space).

### Derivation of entropy constraint equation
We define a graph whose nodes are the 64 basis dimensions of the feature space. (For JPEG this basis is the set of 64 2-d discrete cosine transforms; for RGPEG we derive the generalization of this basis for perceptual space, showing the DCT to be a special case).

Activation patterns in the graph can be thought of as the *state* of the space, and operations on the graph are state transitions. We define $\Omega(x,s)$ as the state describing the intensity of each of the pixels in the (8x8) image, such that $\Omega(x,0)$ is the original image, and any $\Omega(x,s)$ for non-zero *s* values is an altered image, including the possible compressed versions of the image. We define the *s* values to be in units of *bits* x *length*; corresponding to the number of bits required to store a given image, and thus commensurate with entropy (see Supplemental section §2.3). We wish to know how to change the image such that the entropy will be reduced. Changes to the image with respect to entropy are expressed as

$$\frac{\partial \Omega(x,s)}{\partial s}$$

We treat the problem of such image alterations in terms of the heat equation (see, e.g., (27, 28), and see Supplemental section §2.3). We equate the second derivative of the image state with respect to distance, with the derivative of the image with respect to entropy:

$$\frac{\partial \Omega(x,s)}{\partial s} = \frac{\partial^2 \Omega(x,s)}{\partial x^2} \tag{2.7}$$

We term eq. (2.7) the entropy constraint equation; we want to identify $\Psi(x,s)$ that satisfies this equation, such that we can generate modifications of an image to achieve a new image exhibiting a reduction entropy (and correspondingly increase in error).

Via separation of variables we assume a solution of the form
$$\Omega(x,s) = \omega(x)\phi(s) \tag{2.8}$$
where the function $\omega$ is only in terms of position information $x$ and the function $\phi$ only in terms of entropy $s$. Thus the former connotes the "position" portion of the solution, i.e., values of image pixels regardless of entropy values, whereas $\phi$ is the entropy portion of the solution.

We can formulate two ordinary differential equations corresponding to the two sides of the partial differential equation in equation (2.7):

$$\frac{\partial \Omega(x,s)}{\partial s} = \omega(x)\frac{d\phi(s)}{ds} \quad \text{and} \quad \frac{\partial^2 \Omega(x,s)}{\partial x^2} = \frac{d^2\omega(x)}{dx^2}\phi(s)$$

which both equal the same value and can thus be equated:

$$\omega(x)\frac{d\phi(s)}{ds} = \frac{d^2\omega(x)}{dx^2}\phi(s) \tag{2.9}$$

which can be simplified

$$\frac{1}{\phi(s)}\frac{d\phi(s)}{ds} = \frac{1}{\omega(x)}\frac{d^2\omega(x)}{dx^2}$$

Since the two functions are equal, they are equal to some quantity (which cannot be a function of $x$ or $s$, since the equality would then not consistently hold). We call that quantity $\lambda$. There can be a distinct $\lambda$ value for each candidate solution $i$. For any such given solution, the entropy term is:

$$\frac{d\phi(s)}{ds} = \phi_i(s)\lambda_i$$

whose solution is

$$\phi_i(s) = e^{\lambda_i s} \tag{2.10}$$

As mentioned, there will be $i$ solutions for each value of $\lambda$. (See supplemental section §2.3).

For the position term:

$$\frac{d^2\omega(x)}{dx^2} = \omega_i(x)\lambda_i \tag{2.11}$$

the solution is in the form of the Fourier decomposition

$$\omega_i(x) = \sum_i c_i \bar{\gamma}_i(x) \tag{2.12}$$

where the $\bar{\gamma}_i$ terms are the eigenvectors of the Laplacian of the position term, eq. (2.11), and where the $c_i$ terms correspond to the coefficients of the eigenvector basis of the initial condition of the state $\Omega(x,s)$ corresponding to the initial image itself, $\Omega(x,0)$. (Precise formulation of the $c_i$ is shown in the next section). The 64 solutions of the Fourier decomposition form the basis space into which the image will be projected. (For JPEG, this is the discrete cosine transform or DCT set, as mentioned; we will see that this corresponds to one special case of the solution, for a specific set of values of the entropy constraint equation.)

**Application of entropy constraint equation to image feature space**

Consider the graph (Figure 4a) whose nodes are dimensions of feature space $\mathcal{F}$ and whose edges are the pairwise Riemannian distances between those dimensions as defined by the distance matrix of equation (2.6) in section IIIc. The distance matrix can be treated as the adjacency matrix $\bar{A}$ of that graph. We compute the degree matrix $\bar{D}$ via $D_{ii} = \sum_{j=1}^{n} A_{ij}$ for $\bar{A}$ with row indices $i = 1,\ldots,m$ and column indices $j = 1,\ldots,n$. The graph Laplacian is $\bar{L}_g = \bar{D} - \bar{A}$, and the normalized graph Laplacian is then $L = D^{\frac{1}{2}} L_g D^{\frac{1}{2}}$.

The total energy of the system can be expressed in terms of the Hamiltonian $\hat{H}$, taking the form $\hat{H} = L + P$ where $L$ is the Laplacian and $P$ (corresponding to potential energy) can be neglected as a constant for the present case; the hamiltonian is thus equivalent for this purpose to the laplacian:

$$\hat{H} = \frac{\partial^2 \Omega(x,s)}{\partial x^2} \tag{2.13}$$

Intuitively, the Hamiltonian expresses the tradeoffs among different possible states of the system (Figure 4); applied to images, the Hamiltonian can be measured for its errors (distance from the original) on one hand, and its entropy or compactness on the other: a more compact state (lower entropy) will be less exact (higher error), and vice versa.

The aim is to identify an operator that begins with a point in feature space (an image) and moves it to another point such that the changes in error and entropy can be directly measured not in feature space but in perceptual space (Fig 3). Thus the desired operator will move the image from its initial state (with zero "error," since it is the original image, and an initial entropy value corresponding to the information in the image state) to a new state with a new tradeoff between the now-increased error and corresponding entropy decrease.

The Hamiltonian enables formulation of such an operator. The eigenvectors of the Hamiltonian (Figure 4c) constitute a candidate basis set for the image vector (Figure 4d), and since $\hat{H}\Omega = \lambda\Omega$, the eigenvalues $\lambda$ of the Hamiltonian can provide an operator $\hat{U}(s)$ corresponding to any given desired entropy $s$ (see Supplemental section §2.3). As we will see, the $\phi_i(s)$ function is the desired update operator, moving the point corresponding to the state (the image) to a higher-error location in perceptual space to achieve the decreased entropy level corresponding to **s**.

The separated components of the state equation (2.8) form the position solution and entropy solution to the equation, respectively.

$$\Omega(x,s) = \omega(x)\phi(s)$$

The position portion, $\omega(x)$, was shown in equation (2.12) to be

$$\omega_i(x) = \sum_i c_i \bar{\gamma}_i(x)$$

and the entropy portion, $\phi(s)$, was shown in equation (2.10) to be

$$\phi_i(s) = e^{\lambda_i s}$$

The former expresses the set of positional configurations for each given solution and the latter provides the foundation for the update operator for states **i**, to achieve entropy level **s**, where the $\lambda$ values correspond to the eigenvalues of the Hamiltonian.

Combining the terms, we obtain

$$\Omega(x,s) = \omega_i(x)\phi_i(s) = \sum_i c_i \bar{\gamma}_i(x)\bar{\phi}_i(s) \tag{2.14}$$

To put these operations in matrix form, we define the matrix $\bar{\Gamma}$ composed of the column vectors $\bar{\gamma}_i(x)$, i.e., the eigenvectors of equation (2.12). We define the final form of the update operator, $\hat{U}(s)$, to be the matrix composed of column vectors $\bar{\phi}_i(s)$. (Each $\bar{\phi}_i(s)$ has only a single non-zero entry, in the vector location indexed by **i**, and thus $\hat{U}(s)$ is a diagonal matrix).

The transformation steps for altering an image to a degraded image with lowered entropy and increased error, then, begins with the image vector $(\bar{f})$, and projects that vector into the perceptual space defined by the eigenvector basis from equation (2.12), such that

$$\bar{f}' = \Gamma \cdot \bar{f} \tag{2.15}$$

The vector $\bar{f}'$ forms the initial conditions of the original image, transformed into perceptual space (by the $\bar{\Gamma}$ matrix, composed of the $\bar{\gamma}_i$ eigenvectors from equation (2.12) as the columns of $\bar{\Gamma}$). The values $f_i$ of $\bar{f}'$ constitute the values of the $c_i$ coefficients that will be used in equation (2.14).

Having transformed the vector into perceptual space, the update operator is then applied

$$\bar{f}'' = \hat{U}(s) \cdot \bar{f}' = \hat{U}(s) \cdot \bar{\Gamma} \cdot \bar{f} \tag{2.16}$$

The initial image now has been moved into perceptual space $(\bar{f} \to \bar{f}')$, and moved within that space to a point corresponding to entropy level **s** $(\bar{f}' \to \bar{f}'')$, with a corresponding increase in error (which will be measured).

The lower-entropy image, $\bar{f}''$, thus has been scaled such that it now can be encoded via rounding into a more compact version:

$$\bar{f}''' = \text{round}(\bar{f}'') = \text{round}(\hat{U}(s) \cdot \bar{\Gamma} \cdot \bar{f}) \tag{2.17}$$

Any subsequent encoding step may then be applied, such as Huffman or arithmetic coding, operating on the rounded result. These are equivalently applicable to any other method (JPEG, RGPEG, or other) of arriving at a transformed image, and are thus irrelevant to the present formulation. We instead focus on the direct measures of error and of entropy. We proceed to compare these measures directly for JPEG and for the newly introduced RGPEG.

**Update operator moves image to lower entropy state and minimizes error increase**

The image $\bar{f}$ now has been moved from feature space to the perceptual space defined by the eigenvector basis of equation (2.12), as in Figure 4c, selecting a quality level (see Supplemental section §2.3), applying the appropriate update operator, and rounding, resulting in equation (2.17).

As described in section *IIId*, these computations depended on construction of a Jacobian either via knowledge of (or estimated approximation of) the anatomical paths from input to percept (synaptic Jacobian), or via empirical psychophysical measures (psychophysical Jacobian).

We carried out several instances of computed compression via an estimated synaptic Jacobian, composed by measuring distances between pixels on a screen image, measuring viewing distance from the screen, converting these to viewing angle, and measuring all pixels in terms of viewing angles and the distances among them (Supplemental section §2.5, and supplemental table §1). Examples of computed Hamiltonians and eigenvector bases are shown in Figure 4e and 4g for a particular empirical pixel size and viewing distance (Supplemental section §2.5); the formulae show how any empirically measured features give rise to a corresponding Hamiltonian. A set of several additional sample Hamiltonians and eigenvector bases are shown in Supplemental figures §9-§13.

In sum, JPEG assumes its basis vectors (discrete cosine transforms) to be orthogonal, which they are in feature (Euclidean) space, but not in perceptual (Riemannian) space. As shown, the perceptual non-zero distances among basis dimensions can be either empirically ascertained via psychophysical similarity experiments, as in the psychophysical-jacobian method, or assumed on the basis of presumptive measures of anatomical distances (or approximations thereof) as in the synaptic-jacobian method, or calculated on the basis of physically measured distances in the physical space, as in the physical-distance-jacobian method (see Supplemental section §2.5). In the present paper we have predominantly tested the estimated psychophysical jacobian method (method ii b above), which (perhaps surprisingly) is shown, by itself, to outperform JPEG. From these methods, we derived Hamiltonians from the image space, and eigenvector bases from the Hamiltonians, and showed that the JPEG DCT basis was a special case with particular settings shown in Supplemental figure §12.

**Side by side comparison of JPEG / RGPEG**

Performing compression with multiple sets of parameters (see Supplemental Figures §15-§24) yielded empirical results enabling comparisons of the error and entropy measures for the JPEG method and the method (RGPEG) derived from the Riemannian geometric principles described herein.  We have shown that for specific assumptions of geometric distance and of perceived intensity difference, the JPEG method occurs as a special case of the general RGPEG principles (Supplemental section §2.11.1).  It is intriguing to note that, using simple estimations of geometric distance and log scale intensity differences, the generalized RGPEG method typically outperforms the JPEG special case, as expected; Figure 5 shows one such detailed side by side comparison; many more are shown in Supplemental figures §15-§24).  It also is notable that the computational space and time costs for the RGPEG method are identical to those for JPEG (Supplemental section §2.12).

Figure 5 shows a range of compressed versions of a sample image (from the Caltech256 dataset), along with the measures of error (er) and entropy (en) for each image.  The method can most clearly be seen to produce fewer artifacts when compared at relatively high compression levels (high entropy and high error); these are clear to qualitative visual inspection; the figure also shows quantitative plots of the tradeoffs of values among error and entropy for a set of selected quality levels.  Across a range of quality settings, the error and entropy values for RGPEG outperform those for JPEG.

## Discussion: derivation of principles

Of primary interest is not the fact that JPEG compression can readily be outperformed by the generalized RGPEG method; rather, the reason for the outperformance is that RGPEG embodies a novel set of principles of perceptual similarity, and that these principles have explanatory power for the set of perceptual phenomena described (of which JPEG compression is one instance).  We briefly discuss these explanatory principles.

*Physical stimulus similarity is distinct from perceptual stimulus similarity.*
Standard distance measures (Euclidean, Mahalanobis, etc.) (29) do not match human similarity and dissimilarity judgments (e.g., Section IIIc above).  To address this, some standard approaches "re-code" the stimuli to more accurately reflect typical subjects' reported perceived similarity or dissimilarity among stimuli (30, 31).  Yet different individual perceivers can differently register dissimilarity among identical physical stimuli, such as the incompatible similarity judgments of speech sounds by native speakers of different languages (4, 6).  The solution is not to re-code the stimuli, but rather to separately represent physical stimuli (e.g., speech sounds) on one hand, and the particular perceptual mappings of those stimuli on the other, via a metric operation that transforms distances from the reference frame of the physical stimulus space into distances in any given perceiver's perceptual reference frame (Section IIId).

*Perceptual distances are intrinsically Riemannian.*
Euclidean vector distances assume orthogonality of constituent vector dimensions.  This could in theory hold but it is in general not the case for perceptual stimuli.  The constituent dimensions of a vector do not distinguish between "nearby" or "distant" dimensions, but human perceptual judgments typically do.  Riemannian space can intuitively be thought of as having "curved" axes (relative to a tangent space) such that some regions of a given axis are "closer" to some axes and farther from others, quite distinct from Euclidean space.  The tools from differential geometry presented here enable stimuli in Euclidean feature space to be mapped to physical and perceptual spaces; we forward the principle that these mappings underlie judgments of perceptual similarity. This paper focuses on examples in visual domains; additional extensions to auditory stimuli, and to abstract concept categorization are separate findings being pursued.

*Perceptual mappings arise directly from anatomical structure and physiological operation.*
a) A perceptual system cannot "neutrally" process stimuli; any system contains intrinsic assumptions about the relations that occur among the components of any stimuli. A perceptual system connectome encodes a Jacobian either with or without off-diagonal entries, causing it to treat stimulus components (e.g., neighboring pixels in an image) as dependent or independent, respectively, and the nature of any off-diagonal entries determines the exact dependency relations among the components, corresponding to the specific curvature of the metric perceptual space).
b) Cortical neuron receptive fields are often characterized in terms of Gaussians (22-25, 32, 33). Such components produce outputs that compute the partial derivatives of their inputs in just the form needed for the Jacobian and tensor computations posited here; i.e., typical neural assemblies appear tailored to computing transforms into Riemannian target spaces.

*Synaptic plasticity changes the curvature of perceptual space.*
Re-shaping neurons' receptive fields via synaptic modification directly changes the Jacobian mapping and the curvature of the target space. Every synaptic "learning rule" corresponds to a mechanism by which existing metric transforms (arising from the connectome) are modified in response to stimuli. All learning rules can be cast in terms of changing curvature of the projection from input to perceptual space.

*Transforms can be computed from observed behavior.*
Connectomes are almost entirely unmapped in sufficient detail to construct a Jacobian, and in any event perceptual spaces are formed by a combination of successive feedforward stages as well as feedback top-down influences. A given perceiver's perceptual space may nonetheless be elicited empirically by psychophysical measures (section IIId).

*Machine learning is based on the same geometric principles.*
Unsupervised learning rules can readily educe statistical distribution characteristics of data, and typically are judged by measures such as within-category vs. between-category distances (34-36). But the discovery of unsupervised structure is not neutral with respect to metric spaces: in response to a given set of data, different rules cause different changes to the Jacobian, discovering different structure in the data (illustrated in the special case of JPEG encoding, but broadly applicable to learning structure in data). Recent neural net approaches have identified learning methods that can outperform JPEG; the present work, by contrast, outperforms JPEG with no training and no statistics, by identifying previously unnoted fundamentals of perceptual encoding that underlie similarity judgments.

*Further principles arise from study of perceptual transforms.*
In the psychophysical Jacobian method (Section IIId), for instance, perceptual distances arise from the minimum distance within the target Riemannian space, i.e., the geodesic. It could have been the case that other distances might instead have been involved. We forward the principle that perceived distances are predicted by measures of Riemannian minimum distance. Other underlying principles may similarly emerge from further study.

*Application to a well-studied perceptual anomaly.*

Tversky and colleagues (1, 2) showed that perceived similarity judgments of some classes of stimuli violated the triangle inequality: even though stimuli A and B may physically share more features than A and C, the latter may be judged more similar than the former. The present studies suggest that subjects in these experiments are perceiving the stimuli in a Riemannian space (Figure 6), in which a seemingly-direct path from one point to another may entail proceeding via curved Riemannian coördinates, making that (perceived) path longer than alternative paths.

In sum, the new formalism presented here is proposed as a general method for describing and predicting perceptual and cognitive similarity judgments, as a complement to standard vector distance metrics (Euclidean, Mahalanobis, etc.), which are applicable only to measures in non-curved spaces. The results are equally applicable to visual, auditory, and other modalities, as well as to abstract concept data.

At the core of the work are the twin principles that i) sensory stimuli (and arbitrary data) may have internal Riemannian structure, i.e., dependence relations among their (dimensional) component features; and ii) any system, natural or artificial, that processes such data contains intrinsic assumptions or biases about the nature of those dependence relations. Such a system may assume that input data are Euclidean and that their components are thus independent, or the system may assume the presence of any of a very wide variety of inter-component dependencies (such as neighbor or topography relations). We formalize such premises, laying groundwork for extended study of natural perceptual systems and of artificial algorithms for processing, representing, and identifying structure in arbitrary data. Ongoing work is focused on extending the findings to domains beyond vision, with the aim of identifying additional useful applications as well as identifying further fundamental principles of representation.

## Acknowledgements

The authors gratefully acknowledge helpful discussions with Eli Bowen. This research was supported in part by grant N00014-15-1-2132 from the Office of Naval Research and grant N000140-15-1-2823 from the Defense Advanced Research Projects Agency.

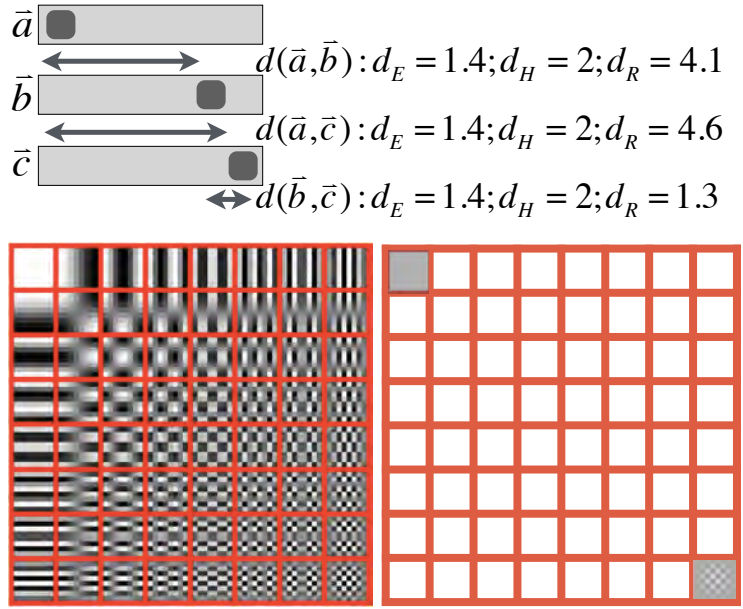

**Figure 1. Illustration of the Riemannian nature of perceptual similarity.** (Top) The transposes of three vectors (1 0 0 0 0 0), (0 0 0 0 1 0), and (0 0 0 0 0 1) ($\vec{a}, \vec{b}, \vec{c}$) are rendered as images with empty space for zeros and dark spots for ones. The Euclidean pairwise distances between any two of $\vec{a}$, $\vec{b}$, and $\vec{c}$ are equal (distances of $\sqrt{2}$). Their Hamming distances also are equal (distances of 2). If we measure the distances between the dark spots, the answers (in mm) come out to be similar from $\vec{a}$ to $\vec{b}$ and from $\vec{a}$ to $\vec{c}$, but quite different (much smaller) from $\vec{b}$ to $\vec{c}$. This "ruler distance" matches the evoked perceptual similarity judgments empirically elicited from human viewers: all judge $\vec{b}$ and $\vec{c}$ to be more similar than either is to $\vec{a}$. (Bottom left) The 64 vectors of the two dimensional discrete cosine transform form an orthogonal basis in Euclidean space; they are equidistant from each other. Perceptual similarity judgments between them, however, exhibit wide variations; some are judged far more similar to each other than others by human perceivers. (Bottom right) Taking just the first and 64th DCT entries (upper left and lower right corners of the DCT, respectively) as an example, when viewed with unit coefficients (as on the left), they are judged quite distinct; however, when viewed with intermediate coefficients they are judged to be somewhat similar (right side). Thus the perceptual metric being used by human viewers apparently is not uniform across this basis space. Thus not only is the space non-Euclidean, it also is non-affine. Throughout this paper, we assume full Riemannian curvature in this basis space.

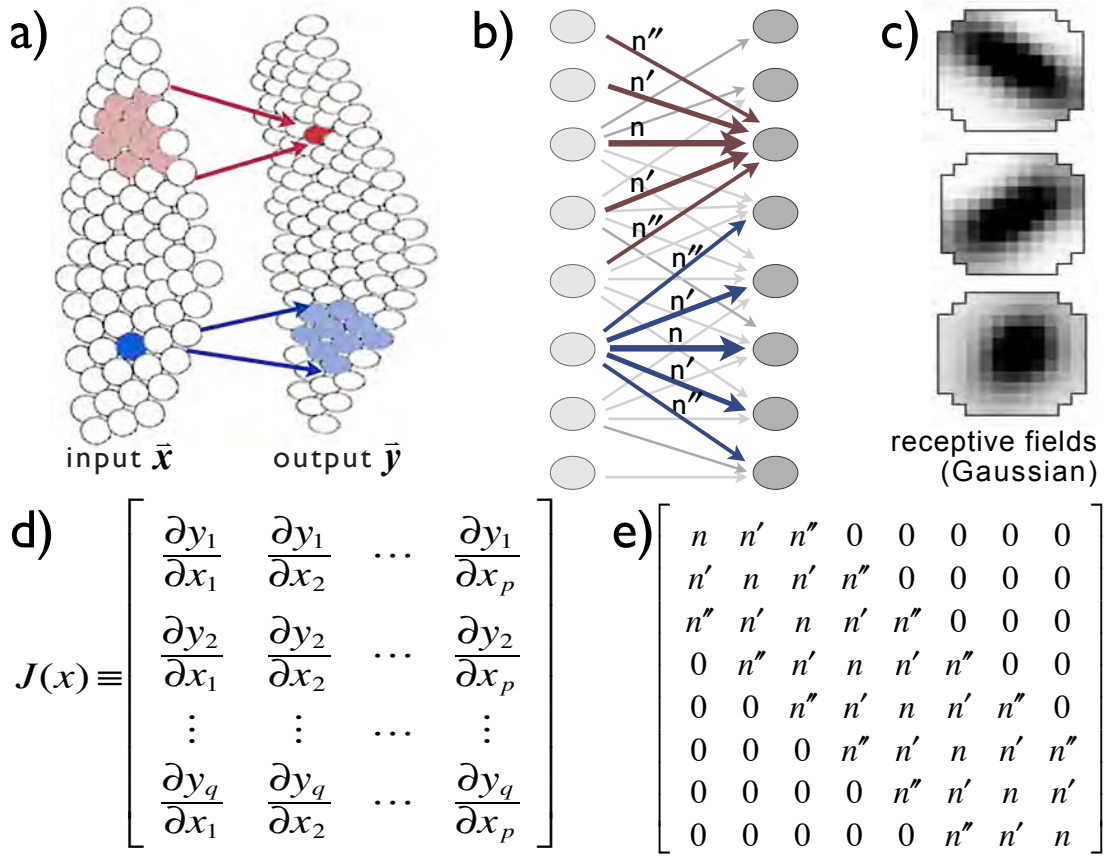

**Figure 2. Brain connectomes are Riemannian.** a) Simple example of anatomical projections between two regions. b) Simple vector encoding of an anatomical projection with synaptic weights. c) Examples of physiological neural responses in early visual areas (gaussians). d) A Jacobian matrix denoting the overall effect of activity in the neurons of an input area ($x$) on the neurons in a target area ($f$); each entry denotes the change in an element of $f$ as consequence of a given change in an element of $x$. e) Example instance of such a Jacobian, corresponding to the synaptic connection pattern in part (b).

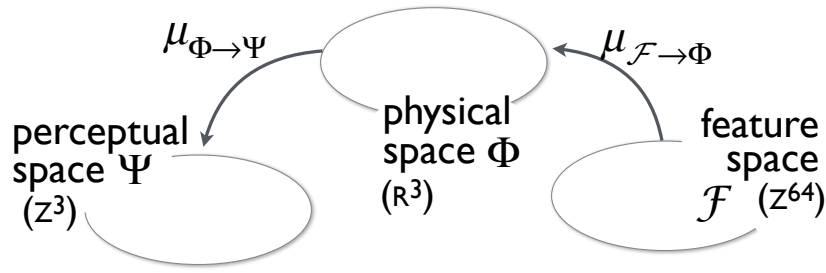

**Figure 3. The map from physical to perceptual space.** The three relevant projection spaces for image compression. For all the examples in this paper, we adopt the JPEG assumption of an 8x8 pixel image. The image consists of a set of intensity settings for each pixel at a given x and y coordinate; this corresponds to Euclidean "physical space" $\Phi$. Images are mapped into feature space, listing the 8x8 pixels as a 64-dimensional vector with integer intensity values from -255 to +255. Human judgments of the similarity of two images (such as an original and a compressed image) correspond to a distinct (Riemannian) space accounting for geometric neighbor relations among the pixels (absent from feature space representation), along with just-noticeable differences (JND) of intensity values at any given pixel. The mapping functions ($\mu$) map from feature to physical space ($\mathcal{F} \rightarrow \Phi$) and from physical to perceptual space ($\Phi \rightarrow \Psi$) as shown.

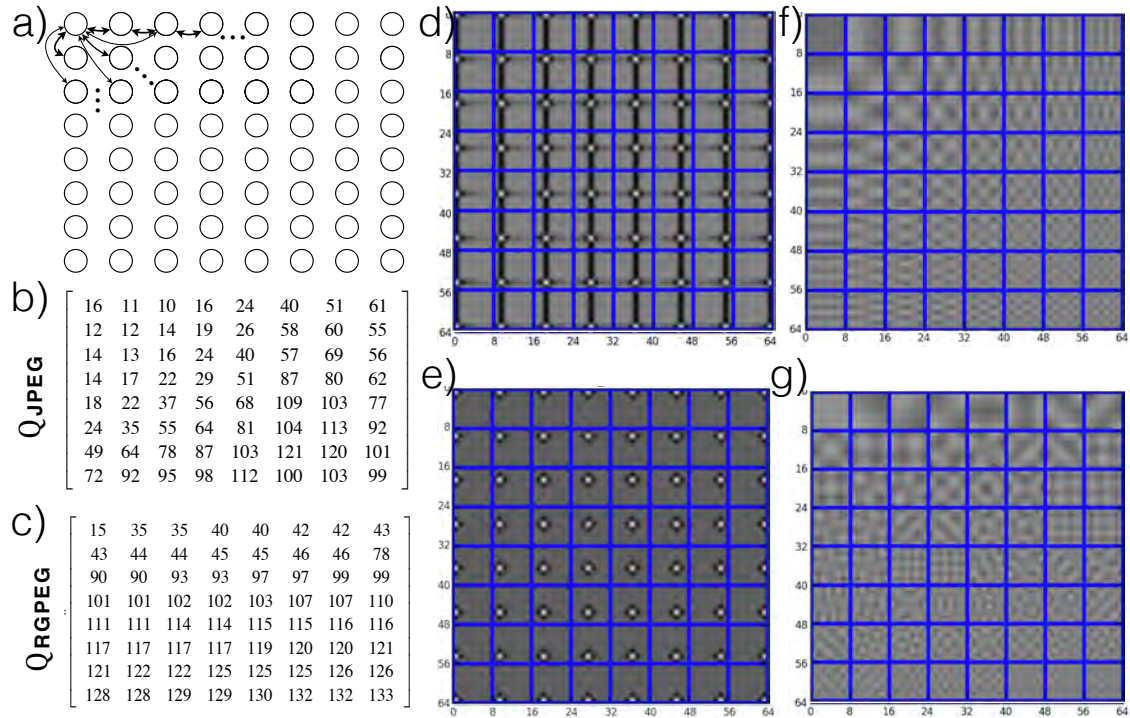

**Figure 4. Treatment of image as graph, and derivation of Hamiltonian.** (a) Basis vectors in feature space $\mathcal{F}$ treated as a graph with whose nodes are the dimensions of the basis and whose edges are the pairwise distances between dimensions (see Eq (2.6)). From that graph, the adjacency and degree matrices, and thus the graph Laplacian, can be directly computed. (b) Q matrix for JPEG (quality level 50%). (c) Computed Q matrix for RGPEG. (d) Hamiltonian for JPEG. (e) Hamiltonian for RGPEG (see Supplemental section §2.7, table §5. (f) Eigenvectors of Hamiltonian for JPEG. (g) Eigenvectors of Hamiltonian for RGPEG. (See Supplemental sections §2.7-2.11).

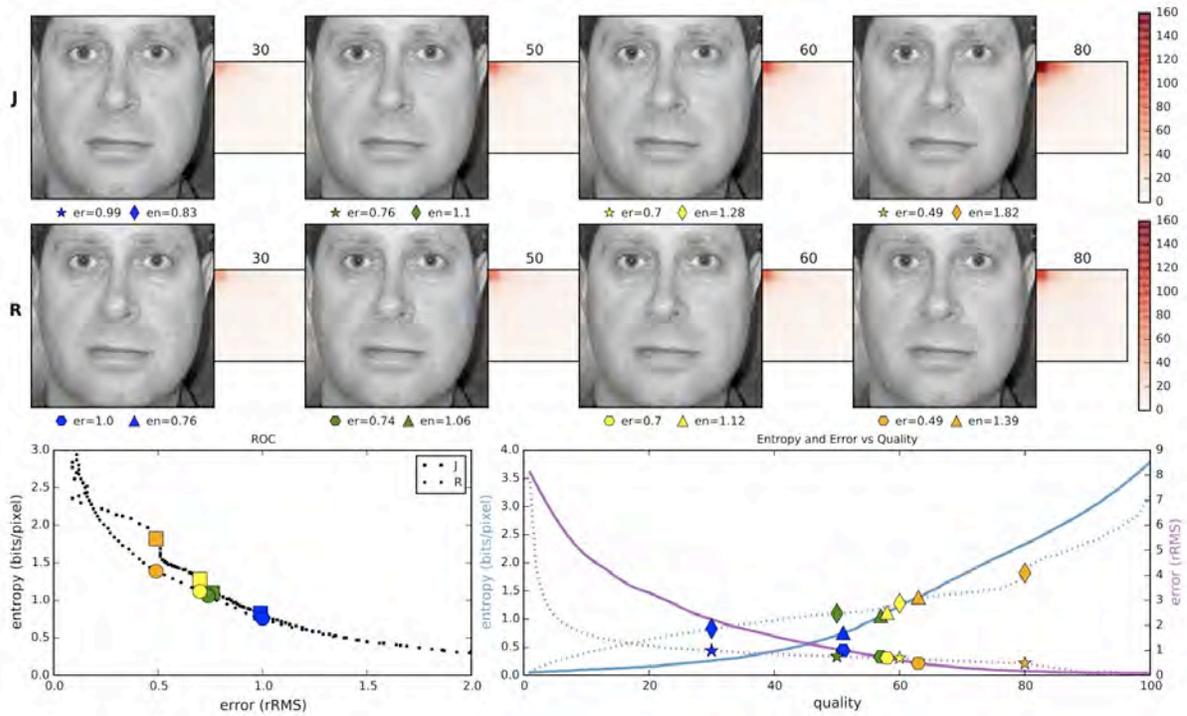

**Figure 5. Side by side comparison of JPEG (J) and RGPEG (R) compression on a sample image.** (For more instances see Supplemental figures §15-§24.) (Top) Examples of images (alongside corresponding computed Jacobians) for given values of desired quality (and corresponding Q matrices), at quality levels 30, 50, 60, and 80, for JPEG (J) and RGPEG (R). For each image, the computed error (er) and entropy (en) are given below the image. For comparable error measures, the entropy for RGPEG is consistently lower than for JPEG. (Bottom left) Receiver operating characteristic for entropy-error tradeoff for JPEG (boxes) and RGPEG (circles). At comparable entropy values, RGPEG error values are consistently equivalent or smaller. (Bottom right) Sample measures of entropy (blue) and error (purple) for JPEG (dotted) and RGPEG (solid) at distinct quality settings. (All images from Caltech-256 (37).

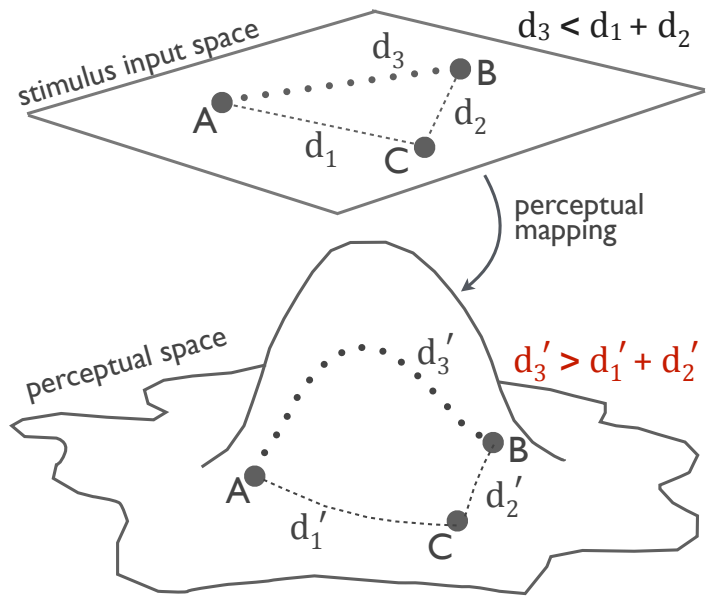

**Figure 6**. Interpretation of the triangle inequality violation (initially described by Tversky and Gati 1982). In a physical stimulus, the distance from A to B is less than the combined distances from A to C to B, i.e., $d_3 \leq d_1 + d_2$, obeying the triangle inequality in the stimulus input space. A perceiver, however, measures those distances not in the input space but in her own perceptual reference frame, which is a Riemannian space (see text). The curvature of that space may render different geodesic distances; specifically, the geodesic from A to B may be longer than the geodesic from A to C to B; thus $d'_3 > d'_1 + d'_2$, violating the triangle inequality in perceptual space.